\documentclass[prl,twocolumn,showpacs]{revtex4}

\usepackage{graphicx}
\usepackage{amsmath}

\begin{document}

\title{Confinement induced molecules in a 1D Fermi gas}

\author{Henning Moritz, Thilo St{\"o}ferle, Kenneth G{\"u}nter, Michael K{\"o}hl$^*$,  and
Tilman Esslinger}

\affiliation{Institute of Quantum Electronics, ETH Z\"{u}rich,
H\"{o}nggerberg, CH--8093 Z\"{u}rich, Switzerland}

\date{\today}

\begin{abstract}

We have observed two-particle bound states of atoms confined in a
one-dimensional matter wave guide. These bound states exist
irrespective of the sign of the scattering length, contrary to the
situation in free space. Using radio-frequency spectroscopy we
have measured the binding energy of these dimers as a function of
the scattering length and confinement and find good agreement with
theory. The strongly interacting one-dimensional Fermi gas which
we create in an optical lattice represents a realization of a
tunable Luttinger liquid.

\end{abstract}


\maketitle

The study of two particles forming a bound state has a long
history both in physics and chemistry because it constitutes the
most elementary chemical reaction. In most situations, such as
atoms in the gas phase or in a liquid, the two particles can be
considered as being in free space and their collisions can be
described by standard quantum mechanical scattering theory. For
ultracold atoms undergoing s-wave interaction a bound molecular
state is only supported when the scattering length between the
atoms is positive whereas for negative scattering length the bound
state is absent \cite{Wigner}.

Weakly bound diatomic molecules in ultracold atomic gases can be
produced using magnetic field induced Fesh\-bach resonances
\cite{Donley2002,Regal2003molecules,Chin2003,Herbig2003,Strecker2003,Xu2003,Cubizolles2003,Jochim2003,Duerr2004}.
The scattering length between the atoms and thus the binding
energy of the molecules can be tuned by an external magnetic
field. The fundamental property that a bound state exists only for
positive scattering length was clearly revealed experimentally
\cite{Regal2003molecules}. For negative scattering length pairing
due to many-body effects has been observed in fermionic atoms
\cite{Chin2004,Greiner2005}. In this work we report on the
observation of bound states of atoms with negative scattering
length where the particles are subject to one-dimensional
confinement. The reduced dimensionality strongly affects the
two-particle physics provided that the scattering length and the
size of the transverse ground state are similar
\cite{Olshanii1998,Petrov2001,Bergemann2003,Mora2004}. This
contrasts with previous studies of interaction-induced phenomena
in one-dimensional quantum systems where the reduced
dimensionality affects only the many-body properties, such as
spin-charge separation in cuprates \cite{Kim1996}, the Mott
insulator transition for bosonic atoms in an optical lattice
\cite{Stoeferle2004}, and the fermionization of a Bose gas
\cite{Tolra2004,Paredes2004,Kinoshita2004}.

Tight transverse confinement alters the scattering properties of
two colliding atoms fundamentally and a bound state exists
irrespective of the sign of the scattering length. This peculiar
behaviour in a one-dimensional system arises from the additional
radial confinement which raises the continuum energy to the zero
point energy of the confining potential, e.g. the two-dimensional
harmonic oscillator ground state energy $\hbar\omega_r$. The
energy of a bound or quasi-bound state remains nearly unaffected
by the external confinement as long as the effective range of the
interaction is small compared to the extension of the confined
ground state. Therefore, a quasi-bound state, which for negative
scattering length $a$ lies above the continuum in free space, is
below the new continuum in the confined system.

The binding energy $E_B$ of dimers in a one-dimensional gas is
given by~\cite{Bergemann2003}
\begin{eqnarray} \label{Ebinding}
\frac{a}{a_{r}}=-\frac{\sqrt{2}}{\zeta(1/2,-E_B/2 \hbar
\omega_{r})},
\end{eqnarray}
where $a_r=\sqrt{\hbar/m \omega_r}$ is the extension of the
transverse ground state (with $m$ being the atomic mass) and
$\zeta$ denotes the Hurwitz zeta function. For negative $a$ and
$|a| \ll a_{r}$ a weakly bound state with $E_B\approx m \omega_r^2
a^2$ exists which has a very anisotropic shape \cite{Tokatly2004}.
In the limit $|a| \gg a_r$ the binding energy takes the universal
form $E_B\approx0.6\,\hbar \omega_r$ and for positive $a$ and
$a\ll a_r$ the usual 3D expression for the binding energy
$E_B=\hbar^2/m a^2$ is recovered.

A trapped gas is kinematically one-dimensional if both the
chemical potential and the temperature are smaller than the level
spacing due to the transverse confinement. For a harmonically
trapped Fermi gas the Fermi energy $E_F=N\cdot \hbar \omega_z$
must be smaller than the energy gap to the first excited state in
the transverse direction $\hbar \omega_r$. $N$ denotes the number
of particles and $\omega_z$ is the trapping frequency along the
weakly confining axis. In our experiment we employ a
two-dimensional optical lattice in order to create 1D Fermi gases.
For atoms trapped in the intensity maxima of the two perpendicular
standing wave laser fields the radial confinement is only a
fraction of the optical lattice period \cite{Greiner2001}. The
much weaker axial trapping is a consequence of the gaussian
intensity envelope of the lattice laser beams. The resulting
aspect ratio $\frac{\omega_r}{\omega_z}=\frac{\pi w}{\lambda}$ is
determined by the waist $w$ and the wavelength $\lambda$ of the
beams. The two-dimensional optical lattice creates an array of 1D
tubes of which approximately $70\times70$ are occupied
\cite{Measured Cloud Size}. This array fulfills the 1D condition
$N<\omega_r/\omega_z \approx 270$ in each tube while
simultaneously providing a good imaging quality.

One-dimensional quantum systems have been realized with fermions,
in e.g. semiconductor nanostructures \cite{Warren1986}, and with
bosons in ultracold atomic gases
\cite{Moritz2003,Stoeferle2004,Tolra2004,Paredes2004,Kinoshita2004}
but so far the scattering length could not be tuned. Here we
overcome this by using a magnetic field induced Feshbach resonance
between two different spin states of the atoms. The resonance is
characterized by its position $B_0$, its width $\Delta B$ and the
background scattering length $a_{bg}$. The scattering length
varies according to $a(B)=a_{bg}(1-\frac{\Delta B}{B-B_0})$ which
allows us to access any value of the scattering length and study
the predicted bound states in 1D.

Our experimental procedure used to produce a degenerate Fermi gas
trapped in an optical lattice has been described in detail in
previous work \cite{Koehl2004}. Fermionic $^{40}$K atoms are
sympathetically cooled by thermal contact with bosonic $^{87}$Rb
atoms, the latter being subjected to forced microwave evaporation.
After reaching quantum degeneracy for both species with typically
$6\times 10^5$ potassium atoms at a temperature of $T/T_F=0.35$
($T_F$ is the Fermi-temperature), we remove all rubidium atoms
from the trap. The potassium atoms are then transferred from the
magnetic trap into a crossed-beam optical dipole trap where we
prepare a spin mixture with $(50 \pm 2)\%$ in each of the $|F=9/2,
m_F=-9/2\rangle$ and $|F=9/2, m_F=-7/2 \rangle$ spin states using
a sequence of two radio-frequency pulses. From now on we will
refer to the atomic state only with its respective $m_F$ number.
By lowering the optical trap depth at a magnetic field of
$B=227$\,G, which is well above the magnetic Feshbach resonance
centered at $B_0=202.1$\,G \cite{Regal2004}, we evaporatively cool
the potassium cloud to a temperature of $T/T_F=0.2$ with
$1.5\times 10^5$ particles.

Prior to loading the atoms into the optical lattice we tune the
magnetic field to $B=210$\,G, so that the s-wave scattering length
between the two states vanishes. The magnetic field strength is
calibrated by radio-frequency spectroscopy between different
Zeeman levels of $^{40}$K and the uncertainty is below 0.1\,G. The
process of loading the atoms from an optical dipole trap made up
from crossing beams along the horizontal x- and y-axes into a
two-dimensional lattice consisting of two standing waves along the
y- and z-axes proceeds as follows: First the standing wave laser
field along the vertical z-axis is turned on. Subsequently, the
optical dipole trap along the y-axis is turned off and a standing
wave laser field along the same axis is turned on. Finally, the
optical trap along the x-axis is turned off. In order to keep the
loading of the atoms into the lattice as adiabatic as possible the
intensities of the lasers are slowly increased (decreased) using
exponential ramps with time constants of 10 ms (25 ms) and
durations of 20 ms (50 ms), respectively. The optical dipole trap
and the lattice are created using the same laser beams which are
focused to 1/$e^2$-radii of $50\,\mu$m (x-axis) and $70\,\mu$m
(y-axis and z-axis) and have a wavelength of $\lambda=826$\,nm.
They possess mutually orthogonal polarizations and their
frequencies are offset with respect to each other by several ten
MHz. The optical potential depth is proportional to the laser
intensity and is conveniently expressed in terms of the recoil
energy $E_r=\hbar^2 k^2/(2m)$, with $k=2 \pi / \lambda$. The
lattice depth was calibrated by intensity modulation and studying
the parametric heating. The calibration error is estimated to be
$<10\%$.


\begin{figure}[htbp]
  \includegraphics[width=.85\columnwidth,clip=true]{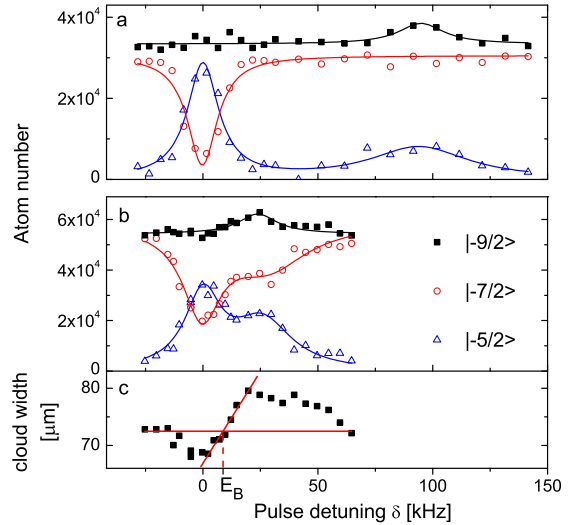}
  \caption{
Radio-frequency (rf) spectroscopy of a one-dimensional gas at
magnetic fields 201.5\,G (a) and 203.1\,G~(b, c) with respective
scattering lengths $2.5\cdot10^3\,a_0$ (a) and
$-1.2\cdot10^3\,a_0$ (b,~c), where $a_0$ is the Bohr radius. The
atom number in the respective spin states is plotted versus the
detuning of the applied rf pulse in (a) and (b). The solid lines
are single or double Lorentzian fits. (c)~Width of the
$|-9/2\rangle$ atom cloud along the 1D tube direction after 7\,ms
time-of-flight obtained from a fit \cite{DeMarco1999} to the
atomic density distribution. The horizontal line marks the average
width for an off-resonant rf pulse, the increase at the molecule
dissociation threshold is fitted using a linear function. The
decrease in width at higher detunings is due to a diminishing
dissociation efficiency.}
  \label{fig1}
\end{figure}

We create molecules by ramping the magnetic field from the zero
crossing of the scattering length at $B=210$\,G in 10 ms to its
desired value close to the Feshbach resonance. Depending on the
final value of this sweep the binding energy of the molecules
varies according to equation (1). We measure the binding energy
$E_B$ of the dimers by radio-frequency spectroscopy
\cite{Regal2003molecules}. A pulse with a frequency $\nu_{RF}$ and
a duration of $40\,\mu s$ dissociates the molecules and transfers
atoms into the initially unpopulated $|-5/2\rangle$ state which
does not exhibit a Feshbach resonance with the state
$|-9/2\rangle$ at this magnetic field. We vary the detuning
$\delta=\nu_{RF}-\nu_{0}$  from the  resonance frequency $\nu_0$
of the atomic $|-7/2\rangle \rightarrow |-5/2\rangle$ transition.
The power and duration of the pulse is optimized to constitute a
$\pi$-pulse on the free atom transition. The number of atoms in
each spin state is detected using absorption imaging after
ballistic expansion. For this we ramp down the lattice
exponentially with a duration of 1\,ms and time constant of
0.5\,ms from the initial depth $V_0$ to $5\,E_r$ to reduce the
kinetic energy of the gas in the transverse directions and then
quickly turn off the trapping potential. The magnetic offset field
is switched off at the start of the expansion, so that no
molecules can be formed in the short time that it passes the
Feshbach resonance. We apply a magnetic field gradient during
3\,ms of the total 7 ms of ballistic expansion to spatially
separate the spin components.

Figure \ref{fig1} shows rf spectra for one-dimensional gases with
a potential depth of the optical lattice of $V_0=25\,E_r$, which
corresponds to $\omega_r=2 \pi \times 69$\,kHz. In
Fig.\;\ref{fig1}a the magnetic field is detuned 0.57\,G below the
Feshbach resonance, i.e. $a>0$. This spectrum exhibits two
resonances: one corresponds to the $|-7/2\rangle \rightarrow
|-5/2\rangle$ transition for free atoms at $\delta=0$, the other
at $\delta>0$ corresponds to dissociated molecules. The
constituent atoms of the dimers are observed in the $|-9/2\rangle$
and $|-5/2\rangle$ states. At this magnetic field, the molecules
are not detected by our state-selective imaging procedure unless
the are dissociated by an rf-pulse. This is due to the fact that
they are transformed into deeply bound molecules during the
switch-off of the magnetic field.

In Fig.\;\ref{fig1}b the magnetic field is chosen 0.95\,G above
the resonance, i.e. $a<0$. Again, the appearance of a second peak
in the $|-5/2\rangle$ atom number at $\delta>0$ demonstrates the
existence of bound state in our 1D geometry. These bound states
are confinement induced since no molecules exist without
confinement above the Feshbach resonance. They are only stabilized
by the presence of the confining potential. Ramping down the
lattice before detection dissociates the dimers and therefore all
atoms should be detected in the image and the total particle
number is expected to remain constant. This is reflected in our
data, where the $|-7/2\rangle$ atom number decreases upon
dissociation while the $|-5/2\rangle$ atom number increases. The
slightly smaller total particle number away from the molecular
resonance could be due to a small fraction of molecules formed
during the switch-off of the magnetic field. Moreover, reduced
losses of the rf dissociated atom pairs with respect to molecules
during the 1\,ms lattice turn off might be held responsible.

The rf pulse not only breaks the pairs if the detuning $\delta$
exceeds the binding energy $E_B$, but also imparts the kinetic
energy $\Delta E= 2 \pi \hbar \delta- E_B$ to the fragments. In
the 1D tubes only the kinetic energy along the tube axis increases
as the motion in the other direction is frozen out for $\Delta E <
\hbar \omega_r$. The cloud width shown in Fig.\;\ref{fig1}c is
extracted from the momentum distribution obtained from time of
flight images. We use this characteristic to determine the binding
energy. This is done by identifying the threshold position at
which the cloud width exceeds that of a cloud without
dissociation. The latter is determined by the Fermi statistics of
the trapped atoms and the interaction of the $|-9/2\rangle$ with
the $|-7/2\rangle$ atoms close to the Feshbach resonance. The
decrease at $\delta \approx 0$ is due to the particle transfer
into the $|-5/2\rangle$ state and an accordingly weaker
interaction energy. Owing to this complication and possible
collisional shifts \cite{Harber2002,Gupta2003} we estimate the
systematic error of our binding energy measurements in all data
sets to be 10 kHz.

\begin{figure}[htbp]
  \includegraphics[width=.85\columnwidth,clip=true]{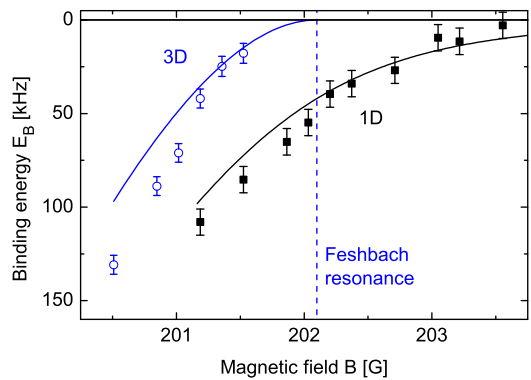}
  \caption{
1D and 3D molecules. Confinement induced molecules in the 1D
geometry exist for arbitrary sign of the scattering length. The
solid lines show the theoretical prediction of the binding energy
with no free parameters (see text). In the 3D case we observed no
bound states at magnetic fields above the Feshbach resonance
(vertical dashed line). The error bars reflect the uncertainty in
determining the position of the dissociation threshold.}
  \label{fig2}
\end{figure}

We have investigated the dependence of the binding energy of the
1D dimers on the magnetic field (Fig.\;\ref{fig2}) and we observed
bound states for every examined magnetic field strength. The
dimers at magnetic fields above the Feshbach resonance are induced
by the confinement. The data is in good agreement with the
theoretical expectation calculated from Eq.\;\ref{Ebinding} (solid
line) with no free parameters. For this calculation we compute the
effective harmonic oscillator length $a_{r}$ and the ground state
energy $\hbar\omega_{r}$ by minimizing the energy of a gaussian
trial wave function in a single well of the lattice to account for
the anharmonicity of the potential. To calculate the scattering
length we use a width of the Feshbach resonance of $\Delta
B=7.8$\,G \cite{Regal2003lifetime} and background scattering
length $a_{bg}=174\,a_0$ \cite{Regal2003background}.

For a comparison with the situation in free space we created
molecules in a crossed beam optical dipole trap without optical
lattice where confinement effects are not relevant. The binding
energy in 3D is measured with the same rf spectroscopy technique
as for the 1D gas and we find molecules only for scattering
lengths $a>0$. The binding energy is calculated according to
\cite{Gribakin1993} as $E_{B,3D}=\frac{\hbar^2}{m (a-\bar{a})^2}$
with $\bar{a}=(m C_6/\hbar^2)^{1/4} \frac{\Gamma(3/4)}{2 \sqrt{2}
\Gamma(5/4)}$ being the effective scattering length and $C_6=3897$
(in atomic units) \cite{Derevianko1999}. The deviation of the
theory from the measured data for more deeply bound molecules is
probably due to limitations of this single channel theory. A
multi-channel calculation would determine the binding energy more
accurately.

\begin{figure}[htbp]
  \includegraphics[width=.85\columnwidth,clip=true]{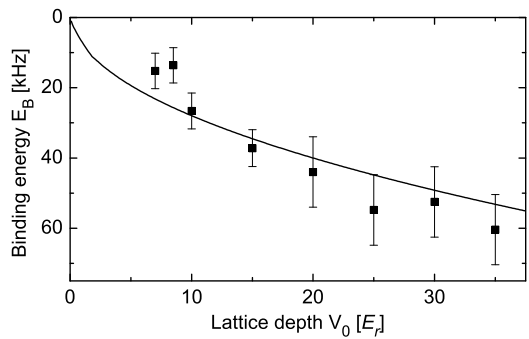}
  \caption{Changing the confinement. The spectra are taken very close to the
  Feshbach resonance at a magnetic field of $B=202.0$\,G. The binding energy
  is measured by rf spectroscopy. For $V_0\geq30\,E_r$ no increase in kinetic energy could be
  detected and we used the rising edge in the $|-5/2\rangle$ atom number in the spectrum to determine the binding
  energy. The error bars reflect the uncertainty in determining the position of the dissociation threshold.
  The solid line shows the theoretically expected value $E_B=0.6 \,\hbar \omega_r$. }
\label{fig3}
\end{figure}

Exactly on the Feshbach resonance where the scattering length
diverges, the binding energy takes the universal form $E_B \approx
0.6 \hbar \omega_r$ and is solely dependent on the external
confinement. We have varied the potential depth of the optical
lattice and thereby the transverse confinement and measured the
binding energy. We find good agreement of our data with the
theoretical prediction (see Fig.\;\ref{fig3}). For a very low
depth of the optical lattice the measured data deviate from the 1D
theory due to the fact that the gas is not one-dimensional
anymore.

In conclusion, we have realized an interacting 1D Fermi gas in a
two-dimensional optical lattice. Using a Feshbach resonance we
have created molecules and measured their binding energy. We find
two-particle bound states for arbitrary sign of the scattering
length, which in the case of negative scattering length are
stabilized only by the tight transverse confinement. We find good
agreement with theory describing the two-particle physics. The
strongly interacting 1D Fermi gas realizes an atomic Luttinger
liquid and fascinating many-body phenomena are predicted in this
system \cite{Recati2003,Astrakharchik2004,Giamarchi2004}.
Especially intriguing appears the BCS-BEC crossover for which
exactly solvable models exist in one dimension
\cite{Tokatly2004,Zwerger2004}.

We would like to thank D. Blume, T. Bergeman, M. Moore, M.
Olshanii, C. Schori, and W. Zwerger for insightful discussions,
and SNF and SEP Information Sciences for funding.


\begin{thebibliography}{}

\bibitem[*]{email}{Email:Koehl@phys.ethz.ch}

\bibitem{Wigner} E. Wigner, Zeitschr. f. Physik {\bf 83}, 253
(1933).

\bibitem{Donley2002} E. A. Donley {\it et al.}, Nature (London) {\bf 417}, 529 (2002).

\bibitem{Regal2003molecules} C. A. Regal, C. Ticknor, J. L. Bohn, and D. S. Jin,
Nature {\bf 424}, 47 (2003).

\bibitem{Chin2003} C. Chin, A. J. Kerman, V. Vuletic, and S. Chu,
Phys. Rev. Lett. {\bf 90}, 033201 (2003)

\bibitem{Herbig2003} J. Herbig {\it et al.}, Science {\bf 301}, 1510 (2003).

\bibitem{Strecker2003} K. E. Strecker,
G. B. Partridge, and R. G. Hulet, Phys. Rev. Lett. {\bf 91},
080406 (2003).

\bibitem{Xu2003} K. Xu {\it et al.}, Phys. Rev. Lett. {\bf 91}, 210402
(2003).

\bibitem{Cubizolles2003} J. Cubizolles, T. Bourdel, S.J.J.M.F. Kokkelmans, G.V. Shlyapnikov, C. Salomon, Phys. Rev. Lett. {\bf
91}, 240401 (2003).


\bibitem{Jochim2003} S. Jochim {\it et al.},
Phys. Rev. Lett. {\bf 91}, 240402 (2003).

\bibitem{Duerr2004} S. D{\"u}rr, T. Volz, A. Marte, and G. Rempe, Phys. Rev. Lett. {\bf 92}, 020406 (2004).

\bibitem{Chin2004} C. Chin {\it et al.}, Science {\bf 305}, 1128
(2004).

\bibitem{Greiner2005} M. Greiner, C. A. Regal, and D. S. Jin,
Phys. Rev. Lett. {\bf 94}, 070403 (2005).


\bibitem{Olshanii1998} M. Olshanii,
Phys. Rev. Lett. {\bf 81}, 938 (1998).

\bibitem{Petrov2001} D. S. Petrov, and G. V. Shlyapnikov, Phys. Rev. A
{\bf 64}, 012706 (2001).

\bibitem{Bergemann2003} T. Bergeman, M. G. Moore, and M. Olshanii,
Phys. Rev. Lett. {\bf 91}, 163201 (2003).

\bibitem{Mora2004} C. Mora, R. Egger, A. O. Gogolin, and A.
Komnik, Phys. Rev. Lett. {\bf 93}, 170403 (2004).

\bibitem{Kim1996} C. Kim {\it et al}, Phys. Rev. Lett. {\bf 77},
4054 (1996).


\bibitem{Stoeferle2004} T. St{\"o}ferle, H. Moritz, C. Schori, M. K{\"o}hl, T.
Esslinger, Phys. Rev. Lett. {\bf 92}, 130403 (2004).

\bibitem{Tolra2004} B. L. Tolra {\it et al.}, Phys. Rev. Lett. {\bf 92}, 190401
(2004).

\bibitem{Paredes2004} B. Paredes {\it et al.},
  Nature {\bf 429}, 227 (2004).

\bibitem{Kinoshita2004} T. Kinoshita, T. Wenger, and D. S. Weiss,
  Science {\bf 305}, 1125 (2004).

\bibitem{Tokatly2004} I. V. Tokatly, Phys. Rev. Lett. {\bf 93}, 090405 (2004).

\bibitem{Greiner2001} M. Greiner, I. Bloch, O. Mandel, T.W. H{\"a}nsch, T. Esslinger,
Phys. Rev. Lett. {\bf 87}, 160405 (2001).

\bibitem{Measured Cloud Size}  For a noninteracting gas in the lattice trap we measure
$1/e^2$ cloud diameters of approximately $60\,\mu$m along the tube
axis and $35\,\mu$m radially.

\bibitem{Warren1986} A. C. Warren, D. A. Antoniadis, and H. I.
Smith, Phys. Rev. Lett. {\bf 56}, 1858 (1986).

\bibitem{Moritz2003} H. Moritz, T. St{\"o}ferle, M. K{\"o}hl, T.
Esslinger, Phys. Rev. Lett. {\bf 91}, 250402 (2003).

\bibitem{Koehl2004} M. K\"ohl, H. Moritz, T.
St\"oferle, K. G\"unter, T. Esslinger, Phys. Rev. Lett. {\bf 94},
080403 (2005).

\bibitem{Regal2004} C. A. Regal, M. Greiner, and D. S. Jin,
Phys. Rev. Lett. {\bf 92}, 040403 (2004).

\bibitem{DeMarco1999} B. DeMarco, and D. S. Jin, Science {\bf 285},
1703 (1999).

\bibitem{Harber2002} D. M. Harber, H. J. Lewandowski, J. M. McGuirk, E. A.
Cornell, Phys. Rev. A {\bf 66}, 053616 (2002);

\bibitem{Gupta2003} S. Gupta {\it et al.}, Science {\bf 300}, 1723 (2003)

\bibitem{Regal2003lifetime} C. A. Regal, M. Greiner, and D. S. Jin,
Phys. Rev. Lett. {\bf 92}, 083201 (2003).

\bibitem{Regal2003background} C. A. Regal, and D. S. Jin, Phys. Rev. Lett. {\bf 90}, 230404 (2003).

\bibitem{Gribakin1993} G. F. Gribakin, and V. V. Flambaum,
  Phys. Rev. A {\bf 48}, 546 (1993).

\bibitem{Derevianko1999} A. Derevianko, W. R. Johnson, M. S. Safronova, J.F. Babb, Phys. Rev. Lett. {\bf 82}, 3589 (1999).

\bibitem{Recati2003} A. Recati, P.O. Fedichev, W. Zwerger, P.
Zoller, Phys. Rev. Lett. {\bf 90}, 020401 (2003); J. Opt. B {\bf
5}, S55 (2003)

\bibitem{Astrakharchik2004} G.E. Astrakharchik, D. Blume, S. Giorgini,
L.P. Pitaevskii, Phys. Rev. Lett. {\bf 93}, 050402 (2004).

\bibitem{Giamarchi2004} T. Giamarchi, {\it Quantum Physics in One
Dimension}, Oxford 2004.

\bibitem{Zwerger2004} J. N. Fuchs, A. Recati, and W. Zwerger, Phys. Rev. Lett. {\bf 93}, 090408 (2004).

\end{thebibliography}
\end{document}